\documentclass{svjour2}                    
\smartqed  
\usepackage{graphicx}
\usepackage{amssymb}
\usepackage{amsmath}
\usepackage{amscd}
\usepackage{mathrsfs}
\usepackage{latexsym}
\usepackage{makeidx}
\usepackage{mathptmx}      
\journalname{Foundations of Physics}
\begin{document}

\title{Generalized Observables,  Bell's Inequalities and Mixtures in the ESR Model for QM
}
\subtitle{A proposal for embedding quantum mechanics into a noncontextual framework}

\titlerunning{ESR model: a proposal for embedding quantum mechanics into a noncontextual framework} 

\author{Claudio Garola         \and
        Sandro Sozzo 
}

\authorrunning{Claudio Garola and Sandro Sozzo} 

\institute{C. Garola \at
              Dipartimento di Fisica dell'Universit\`a del Salento and INFN-Sezione di Lecce, Italy, EU \\
              Tel.: +39-0832-297438\\
              \email{Garola@le.infn.it}           
           \and
           S. Sozzo \at
           Dipartimento di Fisica dell'Universit\`a del Salento and INFN-Sezione di Lecce, Italy, EU \\
              Tel.: +39-0832-297438\\
              \email{Sozzo@le.infn.it}           
}

\date{Received: date / Accepted: date}

\maketitle

\begin{abstract}
The \emph{extended semantic realism} (\emph{ESR}) \emph{model} proposes a new theoretical perspective which embodies the mathematical formalism of standard (Hilbert space) quantum mechanics (QM) into a noncontextual framework, reinterpreting quantum probabilities as \emph{conditional} instead of \emph{absolute}. We provide in this review an overall view on the present status of our research on this topic. We attain in a new, shortened way a mathematical representation of the generalized observables introduced by the ESR model and a generalization of the projection postulate of elementary QM. Basing on these results we prove that the Bell--Clauser--Horne--Shimony--Holt (BCHSH) inequality, a modified BCHSH inequality and quantum predictions hold together in the ESR model because they refer to different parts of the picture of the physical world supplied by the model. Then we show that a new mathematical representation of mixtures must be introduced in the ESR model which does not coincide with the standard representation in QM and avoids some deep problems that arise from the representation of mixtures provided by QM. Finally we get a nontrivial generalization of the L\"{u}ders postulate, which is justified in a special case by introducing a reasonable physical assumption on the evolution of the compound system made up of the measured system and the measuring apparatus. 

\end{abstract}

\section{Introduction\label{intro}}
More than ten years ago one of us, together with another scholar, published some papers aiming to show that a noncontextual (hence local) interpretation of the mathematical formalism of quantum mechanics (QM) was, in principle, possible, contrary to the orthodox view \cite{gs96a,gs96b}. This conclusion was based on the remark that the proofs of the theorems which state that QM necessarily is a contextual and nonlocal theory (mainly the Bell--Kochen--Specker \cite{b66,ks67} and the Bell \cite{b64} theorems) rest on an implicit epistemological assumption about the unrestricted validity of the empirical physical laws of QM (\emph{metatheoretical classical principle}, or \emph{MCP}) that can be questioned from an epistemological point of view and is not consistent with the operational attitude of QM. If one accepts only a weaker assumption (\emph{metatheoretical generalized principle}, or \emph{MGP}) the proofs of the foregoing theorems cannot be completed \cite{ga99,ga99b,ga99c,ga00,ga05,ga08}. The general lines of such a \emph{semantic realism}, or \emph{SR}, \emph{interpretation} were also discussed in the papers quoted above. Of course, the aim of the SR interpretation was to get rid of known quantum paradoxes and avoid the objectification problem \cite{blm91} of the quantum theory of measurement, which find their roots in the contextuality of the orthodox interpretation of QM.

The SR interpretation was, however, rather abstract. To show its consistency some models were propounded \cite{ga02,ga03,gp04}, among which an \emph{extended semantic realism} (\emph{ESR}) \emph{model} that provides a set--theoretical picture of the physical world which preserves the basic feature of the SR interpretation, that is, the substitution of MCP with the weaker principle MGP, but modifies and in some sense extends it. The ESR model is closer to physical intuition and consists of a microscopic and a macroscopic part. The former is a new kind of noncontextual hidden variables theory for QM which introduces, besides hidden variables, a nonstandard interpretation of quantum probabilities, providing a justification of the assumptions introduced in the macroscopic part. The latter can be considered as a new theory that embodies the formalism of QM into a broader noncontextual framework.

As most hidden variables theories, the ESR model presupposes that ``something is happening'' at a microscopic level which underlies the standard quantum picture of the physical world and does not reduce to it (which implies that improvements of the measurements' precision and/or technological developments within the established framework of QM can hardly help in solving the conceptual problems of this theory). One therefore needs a broader theory, and the ESR model aims to be a first step in this direction. According to this model, the macroscopic properties of a given physical system $\Omega$ that can be measured according to QM by macroscopic devices on a \emph{physical object} $x$ (individual example of $\Omega$) bijectively correspond to microscopic properties, each of which either is possessed or is not possessed by $x$, which are the hidden variables of the ESR model (together with further parameters that we do not take into account here for the sake of simplicity and that, however, do not appear in the \emph{deterministic} ESR model \cite{gs08}). The set of all microscopic properties possessed by $x$ is called the \emph{microstate} of $x$. If a macroscopic property $F$ is measured on $x$ and $x$ displays $F$, then $x$ possesses the microscopic property $f$ corresponding to $F$. But the converse implication does not hold, for it can occur that the set of microscopic properties possessed by $x$ is such that $x$ is not detected when $F$ is measured, even if $x$ possesses the property $f$, independently of the specific features of the apparatus measuring $F$. Hence a detection probability is associated with the measurement of $F$ which depends on the microstate of $x$, not only on $f$, and must not be mistook for the detection probability that occurs because of the reduced efficiencies of the real measuring apparatuses.

The introduction of a detection probability depending on the set of all microscopic properties possessed by a physical object is crucial and distinguishes the ESR model from other hidden variables theories in the literature (though some assumptions of the ESR model have been anticipated by several authors when studying particular cases, see, \emph{e.g.}, \cite{sf02}). It implies that one must add a \emph{no--registration outcome} $a_0$ to the set of possible values of any observable $A$ of QM, constructing a \emph{generalized observable} $A_0$ in which $a_0$ is considered as a possible outcome, providing physical information, whenever a measurement of $A_0$ is performed on a physical object $x$ (we stress that $a_0$ occurs also in the case of \emph{idealized} measurements, which correspond to the \emph{ideal first kind} measurements of QM in the ESR model). But the ESR model does not say anything about the deep causes of the detection probability associated with the microstate of the physical object: rather, introduces it as an overall effect of such causes. At a statistical level the introduction implies that detection probabilities occur, to be considered as unknown parameters, whose value is not predicted by any existing theory. The occurrence of such parameters implies some relevant consequences \cite{gs08,g07,s07,g08,s08}, that can be resumed as follows. 

(i) Quantum probabilities can be reinterpreted in such a way that one can recover the mathematical formalism of standard (Hilbert space) QM within the noncontextual (hence local) framework of the ESR model. To be precise, the probability of a property $F$ in a given quantum state $S$ is interpreted as referring to the subset of all physical objects for which the values of the hidden variables (microscopic properties only in the case of a deterministic ESR model) are such that the objects are detected when $F$ is measured, and not to the set of all physical objects that are actually prepared in the state $S$ (in this sense we say that quantum probabilities must be interpreted as \emph{conditional}, not \emph{absolute}).

(ii) Because of noncontextuality, the standard distinction between \emph{actual} and \emph{potential} properties of a physical system in a given state does not occur in the ESR model. The objectification problem of the quantum theory of measurement disappears, together with such paradoxes as ``Schr\"{o}dinger's cat'' and ``Wigner's friend'', because all properties are \emph{objective}: the measurement (or the observer) does not actualize them, and the values of the generalized observables of the physical system can be thought of as assigned for each physical object, independently of any measurement (but, of course, it is impossible to predict all of them even if the quantum state of the object is specified). 

(iii) The detection probabilities introduced by the ESR model can be hardly distinguished from the inefficiencies of real measuring apparatuses, which explains why the former are ignored in QM. But the introduction of these detection probabilities implies also predictions that substantially differ from those of QM and make the ESR model \emph{falsifiable}. In particular, it implies upper limits to the detection inefficiencies in the experiments on Bell's inequalities that can be experimentally checked, at least in principle.   

(iv) More generally, if one reconsiders the physical situation considered by Clauser, Horne, Shimony and Holt to obtain their version of Bell's inequality \cite{chsh69} (briefly, \emph{BCHSH inequality}) from the point of view of the ESR model, one gets a \emph{modified BCHSH inequality}. By introducing some additional assumptions one can then prove that this new inequality may hold together with (suitably reinterpreted) standard quantum results, for it refers to the set of all physical objects that are prepared, while the quantum results refer to the set of all physical objects that can be detected. Thus the known opposition between BCHSH inequality and quantum predictions is overcome, yet in a framework in which physical properties are objective, hence ``local realism'' holds.  

The above conclusions have been achieved without providing an explicit mathematical representation of the new physical entities introduced by the ESR model at a macroscopic level. We have recently supplied such a representation, deducing from it several relevant theoretical consequences. Some of the obtained results have been already published \cite{gs08,g07,s07,g08,s08,gs09,sg08}, some are still in print \cite{gs09b} or unpublished. We propose in this paper a short review on the present status of this part of our research, stressing the contribution that the ESR model can give to solve or avoid some known problems of QM. In detail, the outline of the paper is the following.

We resume in Sect. \ref{modello} the essentials of the ESR model to make the paper self--consistent. Then we supply in Sect. \ref{generalizedformalism} a new proof that each generalized observable must be represented by a family of (commutative) positive operator valued measures, parametrized by the set of all pure states of the physical system $\Omega$, and repropose in Sect. \ref{genprojpost_measproc} a \emph{generalized projection postulate} (GPP), which rules the transformations of pure states induced by idealized nondestructive measurements. These results allow us to prove in Sect. \ref{disgeneralizzate} that the modified BCHSH inequality can coexist with quantum predictions, avoiding the additional assumptions mentioned in (iv). Moreover we show in Sect. \ref{gen_obse_mixed_case} that a new representation of mixtures must be introduced in the ESR model that does not coincide with the standard representation in QM, because each mixture has to be represented by a family of density operators parametrized by the set of all physical properties of $\Omega$. Hence we prove in Sect. \ref{ignorance} that the ESR model establishes a one--to--one correspondence between the operational definition of a mixture and its mathematical representation, which implies that an \emph{ignorance interpretation} of mixtures is possible that avoids some deep problems arising from the representation of mixtures provided by QM. Thus we can propose in Sect. \ref{gpp_mixed_case} a \emph{generalized L\"{u}ders postulate} (GLP) which generalizes GPP in the case of mixtures. Finally we provide in Sect. \ref{justification} a partial dynamical justification of GPP, which introduces nonlinear evolution and avoids the problematic distinction between \emph{proper} and \emph{improper} mixtures.

Let us close this section with a remark. The ESR model recovers noncontextuality (hence locality) by considering only \emph{idealized} measurements. It has been proven by various authors that, whenever actual measurements are considered, contextuality may follow from a statistical description of the experiments on spatially separated systems that adopts a multi--Kolmogorovian rather than a simple Kolmogorovian model as the most natural choice for this class of experiments. This form of contextuality has not a quantum basis and may appear also in classical theories. It occurs, in particular, in the V\"{a}xj\"{o} interpretation of QM \cite{k05a,k05b,k09}, which, as the ESR model, is ``local'' and ``realistic'', and yet predicts experimental violations of Bell's inequalities (a wave model that explains how this breakdown may happen if the measurement apparatuses have thresholds has been recently provided \cite{a09}). In the ESR model a nonproblematic form of contextuality could occur if actual, not only idealized, measurements were considered by introducing additional hidden variables associated with the measuring apparatuses, in agreement with the V\"{a}xj\"{o} view.

\section{The ESR model\label{modello}}
As we have anticipated in Sect. \ref{intro}, we resume in this section the essentials of the ESR model, focusing on the features that are needed in the rest of the paper. More detailed presentations of the model can be found in \cite{ga03,gp04,gs08,g07,g08}.

According to the ESR model, every \emph{physical system} $\Omega$ is characterized at a microscopic level by a set $\mathcal E$ of \emph{microscopic properties} which are in one--to--one correspondence with the macroscopic properties introduced by standard QM, play the role of theoretical entities (\emph{i.e.}, they have no direct physical interpretation) and are such that, for every individual example of $\Omega$ (or \emph{physical object}) $x$, every $f \in \mathcal E$ either is possessed or it is not possessed by $x$, independently of any measurement procedure. Therefore each microscopic property $f$ can be associated with a noncontextual dichotomic \emph{hidden variable}, which takes value $1$ ($0$) if $f$ is possessed (not possessed) by the physical object $x$ that is considered. The set of microscopic properties possessed by $x$ then defines its \emph{microscopic state}, which also plays the role of a theoretical entity. Hence, each microscopic state can be seen as the value of a hidden variable $\lambda$ specifying all microscopic properties of $x$. Whenever a measurement of a macroscopic property is performed on a physical object $x$, the microscopic state of $x$ determines a probability (which is either 0 or 1 if the ESR model is \emph{deterministic}) that the macroscopic apparatus does not react, or, equivalently, that the apparatus remains in its \emph{ready state} and its pointer does not move from its initial position $a_0$. But, then, $a_0$ can be considered as a further possible outcome which provides a peculiar information about $x$, because it informs us that the set of the values of the hidden variables (microscopic properties only if the ESR model is deterministic) is such that $x$ cannot be detected. This interpretation of $a_0$ suggests that a \emph{no--registration outcome} must be added to the set of all possible outcomes of any macroscopic observable. Hence we characterize the physical system $\Omega$ at a macroscopic level by means of a conventional set $\mathcal S$ of \emph{macroscopic states} and a new set $\mathcal O$ of \emph{generalized observables}. Each state $S \in {\mathcal S}$ is operationally defined as a class of physically equivalent \emph{preparing devices} \cite{bc81} which are such that every preparing device $\pi \in S$, when constructed and activated, performs a preparation of a physical object $x$ (we briefly say that ``$x$ is in the state $S$'' in this case). Each generalized observable $A_0 \in {\mathcal O}$ is operationally defined as a class of physically equivalent \emph{measuring apparatuses}, and it is obtained in the ESR model by considering an observable $A$ of QM with set of possible outcomes $\Xi$ on the real line $\Re$ and adding a further outcome $a_0 \in \Re \setminus \Xi$ (\emph{no--registration outcome} of $A_0$), so that the set of all possible values of $A_0$ is $\Xi_0=\{ a_0 \} \cup \Xi$.\footnote{One assumes here, for the sake of simplicity, that $\Re \setminus \Xi$ is non--void. This assumption is not restrictive. Indeed, if $\Xi=\Re$, one can choose a bijective Borel function $f: \Re \rightarrow \Xi'$ such that $\Xi' \subset \Re$ (\emph{e.g.}, $\Xi'=\Re^{+}$) and replace $A$ by $f(A)$. \label{borel_sets}}

Let now $\mathbb{B}(\Re)$ be the $\sigma$--algebra of all Borel subsets of $\Re$. The set 
\begin{equation}
{\mathcal F}_{0} \ =  \ \{ (A_0, X) \ | \  A_0 \in {\mathcal O}, \ X \in \mathbb{B}(\Re) \}
\end{equation}
is the set of all \emph{macroscopic properties} of $\Omega$, for any pair $(A_0, X)$ is interpreted as the property that the value of $A_0$ belongs to $X$. Hence the subset of all macroscopic properties associated with observables of QM (which bijectively corresponds to the set $\mathcal E$ of all microscopic properties, as we have assumed above) is
\begin{equation}
{\mathcal F} \ =  \ \{ (A_0, X) \ | \  A_0 \in {\mathcal O}, \ X \in \mathbb{B}(\Re), \ a_0 \notin X \}.
\end{equation}
 
A measurement of a macroscopic property $F=(A_0, X)$ on a physical object $x$ in the state $S$ is then described as a \emph{registration} performed by means of a \emph{dichotomic registering device} (which may be constructed by using one of the apparatuses associated with $A_0$) whose outcomes are denoted by \emph{yes} and \emph{no}. The measurement yields outcome yes/no (equivalently, $x$ \emph{displays}/\emph{does not display} $F$) if and only if the value of $A_0$ belongs/does not belong to $X$. Whenever $F=(A_0, X) \in \mathcal F$ (hence $a_0 \notin X$) the overall probability $p_{S}^{t}(F)$ that a physical object $x$ in the state $S$ display $F$ when $F$ is measured on $x$ is given by
\begin{equation} \label{formuladipartenza}
p_{S}^{t}(F)= p_{S}^{d}(F)p_{S}(F) \ .
\end{equation}
The symbol $p_{S}^{d}(F)$ in Eq. (\ref{formuladipartenza}) denotes the probability that $x$ be detected whenever $x$ is in the state $S$ (\emph{detection probability}) and $F$ is measured, and it is not necessarily fixed for a given observable $A_0$ but it may depend on the macroscopic property $F$, hence on the Borel set $X$. We assume in the following that, for every $F \in {\mathcal F}$, an \emph{idealized} measurement exists such that $p_{S}^{d}(F)$ depends only on the features of the physical objects in the state $S$, hence it does not occur because of inefficiences of the apparatus measuring $F$, and consider only measurements of this kind.\footnote{This assumption can be justified by considering the microscopic part of the ESR model (see Sect. \ref{intro}). We do not insist on this topic for the sake of brevity. The interested reader can refer to \cite{ga03,gp04,gs08,g07,g08}.\label{intrinsic_feat}} The symbol $p_{S}(F)$ in Eq. (\ref{formuladipartenza}) denotes instead the probability that $x$ display $F$ when it is detected. The following assumption is then basic in the ESR model.

\vspace{.1cm}
\noindent
\emph{AX. If $S$ is a pure state the probability $p_{S}(F)$ can be evaluated by using the same rules that yield the probability of $F$ in the state $S$ according to QM.}

\vspace{.1cm} 
Assumption AX allows one to recover the formalism of QM in the framework of the ESR model, but modifies the standard interpretation of quantum probabilities. Indeed, according to QM, whenever an ideal measurement of a property $F$ is performed, all physical objects that are prepared in a state $S$ are detected. The quantum rules for calculating probabilities are thus intuitively interpreted as yielding the probability that a physical object $x$ display the property $F$ whenever it is selected in the set of all objects in the state $S$, and in this sense we say that they provide \emph{absolute} probabilities in QM. According to assumption AX, instead, if $S$ is pure, the same rules yield the probability that a physical object $x$ display the property $F$ whenever it is selected in the subset of all objects in the state $S$ that can be detected, and in this sense we say that they provide \emph{conditional} probabilities in the ESR model. This reinterpretation implies that the predictions of the ESR model may be different from those of QM, even if the formalism of QM is embodied in the model.

To complete our discussion, let us now consider a macroscopic property $G=(A_0, Y) \in {\mathcal F}_{0} \setminus {\mathcal F}$, hence $a_0 \in Y$, and put $X= Y \setminus \{ a_0 \}$, $F=(A_0, X)$, $F^{c}=(A_0, \Re \setminus Y)$. Then, for every state $S$, we get from Eq. (\ref{formuladipartenza})
\begin{equation} \label{formuladipartenza_0}
p_{S}^{t}(G)=1-p_{S}^{t}(F^{c})=1-p_{S}^{d}(F^{c})p_{S}(F^{c}) ,
\end{equation}
where $p_{S}^{t}(G)$ obviously denotes the overall probability that a physical object $x$ in the state $S$ display $G$ when $G$ is measured on $x$. Let us assume further that $p_{S}^{d}({F}^{c})=p_{S}^{d}(F)$, which is physically reasonable because ${F}^{c}$ can be measured by the same dichotomic registering device measuring $F$, and note that, obviously, $p_{S}({F}^{c})=1-p_{S}(F)$. Then, we get
\begin{equation} \label{f_bar_c}
p_{S}^{t}(G)=1-p_{S}^{d}(F)(1-p_{S}(F))=1-p_{S}^{d}(F)+p_{S}^{t}(F),
\end{equation}
which provides the overall probability of a property in ${\mathcal F}_{0} \setminus {\mathcal F}$ in terms of the overall probability of a property in $\mathcal F$. Moreover, we get from Eqs. (\ref{formuladipartenza_0}) and (\ref{f_bar_c}), by introducing the probability $p_{S}^{t,F}((A_0, \{ a_0 \}))=1-p_{S}^{d}(F)$ that $x$ be not detected when $F$ is measured on it, 
\begin{equation} \label{normalization}
p_{S}^{t,F}((A_0, \{ a_0 \}))+p_{S}^{t}(F)+p_{S}^{t}({F}^{c})=1,
\end{equation}
which expresses a fundamental result that is generalized in the following section.

\section{The mathematical representation of generalized observables\label{generalizedformalism}}
Because of assumption AX, if $S$ is a pure state the probability $p_{S}(F)$ in Eq. (\ref{formuladipartenza}) can be evaluated by using the formalism of QM. Therefore, as far as $p_{S}(F)$ is concerned, the physical system $\Omega$ can be associated with a (separable) complex Hilbert space $\mathscr H$, every pure state $S$ of $\Omega$ can be represented by a unit vector $|\psi\rangle \in \mathscr H$ or by a one--dimensional (orthogonal) projection operator $\rho_{\psi}=|\psi\rangle\langle\psi|$ on $\mathscr H$, and every $F \in \mathcal F$ can be represented by an (orthogonal) projection operator on $\mathscr H$. The probability $p_{S}^{d}(F)$ (hence $p_{S}^{t}(F)$), instead, cannot be obtained by using quantum rules, and we have as yet no theory which allow us to predict it (but, of course, one can try to contrive experiments to determine it empirically). Nevertheless, we have provided a mathematical expression for $p_{S}^{t}(F)$, hence a mathematical representation of the generalized observables introduced by the ESR model, by considering $p_{S}^{d}(F)$ as an unknown parameter in \cite{s08,gs09,sg08}. We intend to provide a new, synthetic approach to this topic in the present section.

Let $A$ be an observable of QM, let $\Xi \subset \Re$ (see footnote \ref{borel_sets}) be the set of its possible outcomes and let $A_0$ be the generalized observable obtained from $A$, whose set of possible outcomes is $\{ a_0 \} \cup \Xi$. We denote by $\widehat{A}$ the self--adjoint operator representing $A$ (the spectrum of which obviously coincides with $\Xi$) and by $P^{\widehat{A}}$ the projection valued (PV) measure associated with $\widehat{A}$ by the spectral theorem,
\begin{equation}
P^{\widehat{A}}: X \in \mathbb{B}(\Re) \longmapsto P^{\widehat{A}}(X) \in {\mathscr L}({\mathscr H}),
\end{equation}
where ${\mathscr L}({\mathscr H})$ is the set of all orthogonal projection operators on ${\mathscr H}$ (hence 
$\widehat{A}=\int_{\Re} \lambda \mathrm{d} P^{\widehat{A}}_{\lambda}$, $\int_{\Re} \mathrm{d} P^{\widehat{A}}_{\lambda}=I$, and, for every $X \in \mathbb{B}(\Re)$,
$P^{\widehat{A}}(X)= \int_{X} \mathrm{d} P^{\widehat{A}}_{\lambda}$). Measuring $A_0$ is then equivalent to measuring all macroscopic properties of the form $F=(A_0, X)$, with $X \in {\mathbb B}(\Re)$, simultaneously. In particular, if one considers an interval $\mathrm{d} \lambda$, with $a_0 \notin \mathrm{d} \lambda$, and the infinitesimal overall probability $\mathrm{d} p_{S}^{t}$ that an idealized measurement of $A_0$ on a physical object $x$ in a pure state $S$ represented by the unit vector $|\psi\rangle \in \mathscr H$ yield an outcome in $\mathrm{d} \lambda$, Eq. (\ref{formuladipartenza}) and assumption AX in Sect. \ref{modello} suggest that
\begin{equation} \label{overall_infinitesimal}
\mathrm{d} p_{S}^{t}=p_{\psi}^{d}(\widehat{A}, \lambda)\langle\psi|\mathrm{d} P^{\widehat{A}}_{\lambda}|\psi\rangle , 
\end{equation}
where $p_{\psi}^{d}(\widehat{A}, \lambda)$ is a detection probability such that $\langle\psi|p_{\psi}^{d}(\widehat{A}, \lambda) \frac{d P_{\lambda}^{\widehat{A}}}{d \lambda}{|\psi\rangle}$ is a measurable function on $\Re$. If $X \in {\mathbb B}(\Re)$ and $a_0 \notin X$, Eq. (\ref{overall_infinitesimal}) implies
\begin{equation} \label{prob_dis_lebesgue_not}
p_{S}^{t}((A_0, X))=\langle\psi | \int_{X} p_{\psi}^{d}(\widehat{A}, \lambda) \mathrm{d} P^{\widehat{A}}_{\lambda}|\psi\rangle .
\end{equation}
Furthermore, Eq. (\ref{normalization}) can now be generalized as follows,
\begin{equation} \label{normalization_lebesgue}
p_{S}^{t}((A_0, \{ a_0 \}))+\langle\psi | \int_{\Re} p_{\psi}^{d}(\widehat{A}, \lambda) \mathrm{d} P^{\widehat{A}}_{\lambda} |\psi\rangle=1
\end{equation}
where $p_{S}^{t}((A_0, \{ a_0 \}))$ is the overall probability that the measurement of $A_0$ yield the $a_0$ outcome. Hence, if $X \in {\mathbb B}(\Re)$ and  $a_0 \in X$, we get
\begin{equation}
p_{S}^{t}((A_0, X))=p_{S}^{t}((A_0, \{ a_0 \}))+p_{S}^{t}((A_0, X \setminus \{ a_0 \})) .
\end{equation} 
Since $a_0 \notin \Xi$, we obtain, by using Eqs. (\ref{prob_dis_lebesgue_not}) and (\ref{normalization_lebesgue})
\begin{equation} \label{prob_dis_lebesgue_yes}
p_{S}^{t}((A_0, X))=\langle\psi |(I- \int_{\Re \setminus X} p_{\psi}^{d}(\widehat{A}, \lambda) \mathrm{d} P^{\widehat{A}}_{\lambda})|\psi\rangle .
\end{equation}
Putting together Eqs. (\ref{prob_dis_lebesgue_not}) and (\ref{prob_dis_lebesgue_yes}) we see that $p_{S}^{t}((A_0, \cdot))$ is a probability measure on the $\sigma$--algebra ${\mathbb B}(\Re)$ of all Borel subsets of $\Re$.

Because of Eqs. (\ref{prob_dis_lebesgue_not}) and (\ref{prob_dis_lebesgue_yes}) one can introduce, for every unit vector $|\psi \rangle \in \mathscr H$, a mapping
\begin{equation}
T_{\psi}^{\widehat{A}}: X \in \mathbb{B}(\Re) \longmapsto T_{\psi}^{\widehat{A}}(X) \in {\mathscr B}({\mathscr H})
\end{equation}
defined by setting
\begin{equation} \label{POV_lebesgue}
T_{\psi}^{\widehat{A}}(X) = \left \{
\begin{array}{cll} 
\int_{X}{p}_{\psi}^{d}(\widehat{A}, \lambda) \mathrm{d} P^{\widehat{A}}_{\lambda} & & \textrm{if} \ a_0 \notin X  \\
I - \int_{\Re \setminus X}{p}_{\psi}^{d}(\widehat{A}, \lambda) \mathrm{d} P^{\widehat{A}}_{\lambda}  & & \textrm{if} \ a_0 \in X
\end{array}
\right. .
\end{equation}
It follows at once from Eq. (\ref{POV_lebesgue}) that $T_{\psi}^{\widehat{A}}$ is a POV measure on $\Re$ which is \emph{commutative}, \emph{i.e.}, for every $X,Y \in \mathbb{B}(\Re)$, $T_{\psi}^{\widehat{A}}(X)T_{\psi}^{\widehat{A}}(Y)=T_{\psi}^{\widehat{A}}(Y)T_{\psi}^{\widehat{A}}(X)$ \cite{blm91}. Hence the discrete generalized observable $A_0$ can be represented by the \emph{family of commutative POV measures}
\begin{equation}
\left \{ T_{\psi}^{\widehat{A}}: X \in \mathbb{B}(\Re) \longmapsto T_{\psi}^{\widehat{A}}(X) \in {\mathscr B}({\mathscr H}) \right \}_{\Vert |\psi\rangle\Vert=1} \ .
\end{equation}
Indeed, bearing in mind Eqs. (\ref{prob_dis_lebesgue_not}) and (\ref{prob_dis_lebesgue_yes}), one gets that the probability that the outcome of a measurement of $A_0$ on a physical object $x$ in the pure state $S$ represented by the unit vector $|\psi\rangle$ lie in the Borel set $X$ is given by
\begin{equation} \label{prob_X_psi}
p_{S}^{t}((A_0, X))=\langle\psi|T_{\psi}^{\widehat{A}}(X)|\psi\rangle .
\end{equation}
Equivalently, one gets
\begin{equation} \label{prob_X_W}
p_{S}^{t}((A_0, X))=Tr[\rho_{\psi}T_{\psi}^{\widehat{A}}(X)] 
\end{equation}
if $S$ is represented by the one--dimensional projection operator $\rho_{\psi}$.

We have thus obtained a mathematical representation of the generalized observables introduced by the ESR model, as desired. To relate this representation with the results resumed in Sect. \ref{modello} let us consider the property $F=(A_0, X)$, with $a_0 \notin X$. We get from Eq. (\ref{formuladipartenza}) and assumption AX
\begin{equation} \label{f_bar}
p_{S}^{t}(F)=p_{S}^{d}(F) \langle \psi | \int_{X} \mathrm{d} P^{\widehat{A}}_{\lambda}|\psi\rangle,
\end{equation}
while Eqs. (\ref{POV_lebesgue}) and (\ref{prob_X_psi}) yield
\begin{equation} \label{f_bar_math}
p_{S}^{t}(F)= \langle \psi | \int_{X} p_{\psi}^{d}(\widehat{A}, {\lambda})\mathrm{d} P^{\widehat{A}}_{\lambda}|\psi\rangle \ ,
\end{equation}
hence
\begin{equation} \label{prova_trattazione_nuova}
p_{S}^{d}(F)=\frac{\langle \psi | \int_{X} p_{\psi}^{d}(\widehat{A},{\lambda}) \mathrm{d} P^{\widehat{A}}_{\lambda} |\psi\rangle}{\langle \psi | \int_{X} \mathrm{d} P^{\widehat{A}}_{\lambda} |\psi\rangle} ,
\end{equation}
or, equivalently, 
\begin{equation} \label{prova_trattazione_nuova_rho}
p_{S}^{d}(F)=\frac{Tr[\rho_{\psi}T_{\psi}^{\widehat{A}}(X)]}{Tr[\rho_{\psi}P^{\widehat{A}}(X)]} .
\end{equation}
Eq. (\ref{prova_trattazione_nuova}) (equivalently, Eq. (\ref{prova_trattazione_nuova_rho})) establishes a relation among detection probabilities which is a direct consequence of the assumption expressed by Eq. (\ref{normalization_lebesgue}). We note explicitly that it entails, for every $|\psi\rangle \in {\mathscr H}$,
\begin{equation} \label{prova_trattazione_nuova_2}
\langle \psi |  \int_{X} (p_{S}^{d}(F)-p_{\psi}^{d}(\widehat{A},{\lambda})) \mathrm{d} P^{\widehat{A}}_{\lambda} |\psi\rangle=0,
\end{equation}
which does not imply $p_{S}^{d}(F)-p_{\psi}^{d}(\widehat{A}, \lambda)=0$, because $p_{S}^{d}(F)-p_{\psi}^{d}(\widehat{A}, \lambda)$ generally depends on $|\psi\rangle$.

Let us compare now the representation of generalized observables introduced here with the representation of observables introduced by unsharp QM \cite{blm91,bgl96,bl96}. Two basic differences spring out.

(i) A generalized observable is represented by a family of POV measures parametrized by the set of all vectors representing pure states, while an observable of unsharp QM is represented by a single POV measure.

(ii) Only commutative POV measures appear in the representation of a generalized observable.

Difference (i) is relevant since it makes explicit that the generalized observables introduced by the ESR model do not coincide, in general, with the observables  introduced by unsharp QM. This can be intuitively explained by recalling that the occurrence of the no--registration outcome, hence of the detection probabilities, is assumed to depend on intrinsic features of the physical object that is considered, while it neither depends on the measuring apparatus (Sect. \ref{modello}) nor it has an unsharp source. Of course this assumption is introduced to recover objectivity of macroscopic properties, avoiding the objectification problem which remains unsolved in unsharp QM \cite{bs96,b98}.

Difference (ii) is less relevant, because only idealized measurements are considered in the ESR model, which correspond to sharp measurements in unsharp QM. It is then reasonable to think that an unsharp extension of the ESR model could be provided by introducing unsharp generalized observables represented by families of noncommutative POV measures. 

Finally, let us illustrate our results by considering a special case.

Let $A$ be a discrete observable of QM, let $\Xi= \{ a_1, a_2, \ldots \}$ be the set of all its possible outcomes, and let $A_0$ be a generalized observable obtained from $A$, with set of possible outcomes $\Xi_0= \{ a_0 \} \cup \{ a_1, a_2, \ldots \}$. We denote by $\widehat{A}$ the self--adjoint operator representing $A$, and by $P_{1}^{\widehat{A}}$, $P_{2}^{\widehat{A}}$, \ldots the (orthogonal) projection operators associated with $a_1$, $a_2$, \ldots, respectively, by the spectral decomposition of $\widehat{A}$. We also put, for every $n \in {\mathbb N}$, $p_{\psi n}^{d}(\widehat{A})\equiv p_{\psi}^{d}(\widehat{A}, a_n)$. Then we get from Eq. (\ref{POV_lebesgue})
\begin{equation} \label{caso_particolare_X}
T_{\psi}^{\widehat{A}}(X) = \left \{
\begin{array}{cll} 
\sum_{n, a_n \in X }p_{\psi n}^{d}(\widehat{A})P_{n}^{\widehat{A}} & & \textrm{if} \ a_0 \notin X  \\
I -\sum_{n, a_n \in \Re \setminus X} (p_{\psi n}^{d}(\widehat{A}))P_{n}^{\widehat{A}} & & \textrm{if} \ a_0 \in X
\end{array}
\right. .
\end{equation}
Let $X=\{ a_n \}$, with $n \in {\mathbb N}_{0}$. Then Eq. (\ref{caso_particolare_X}) yields
\begin{equation} \label{caso_particolare_uno}
T_{\psi}^{\widehat{A}}(\{ a_n \}) = \left \{
\begin{array}{cll} 
p_{\psi n}^{d}(\widehat{A})P_{n}^{\widehat{A}} & & \textrm{if} \ n \ne 0  \\
\sum_{m \in {\mathbb N}} (1-p_{\psi m}^{d}(\widehat{A}))P_{m}^{\widehat{A}} & & \textrm{if} \ n=0
\end{array}
\right. .
\end{equation}
Furthermore, if we put $F_n=(A_0, \{ a_n \})$ Eq. (\ref{prob_X_psi}) yields
\begin{equation} \label{caso_particolare_n}
p_{S}^{t}(F_n)=\left \{
\begin{array}{cll}
p_{\psi n}^{d}(\widehat{A}) \langle\psi|P_{n}^{\widehat{A}}|\psi\rangle & &  \textrm{if} \ n \ne 0 \\
\sum_{m \in {\mathbb N}} (1-p_{\psi m}^{d}(\widehat{A})) \langle\psi|P_{m}^{\widehat{A}}|\psi\rangle & & \textrm{if} \ n=0
\end{array}
\right. .
\end{equation}

\section{The generalized projection postulate\label{genprojpost_measproc}}
The mathematical representation of generalized observables provided in Sect. \ref{generalizedformalism} leads one to inquire into the state transformation induced by measurements of physical properties. If one considers a \emph{nondestructive} idealized measurement, consistency with assumption AX suggests that, if the state $S$ of a physical object $x$ is pure and a sharp value of a discrete observable is obtained, then $S$ is modified according to standard QM rules whenever $x$ is detected. This requirement, together with the results obtained in Sect. \ref{generalizedformalism}, supports the introduction of the following \emph{generalized projection postulate}.

\vspace{.1cm}
\noindent
\emph{GPP}. \emph{Let $S$ be a pure state represented by the unit vector $|\psi\rangle$ or, equivalently, by the density operator $\rho_{\psi}=|\psi\rangle\langle\psi|$, and let a nondestructive idealized measurement of a physical property $F=(A_0,X)\in {\mathcal F}_{0}$ be performed on a physical object $x$ in the state $S$. Let the measurement yield the yes outcome. Then, the state $S_F$ of $x$ after the measurement is a pure state represented by the unit vector}
\begin{equation} \label{genpost_dis_psi}
|\psi_{F} \rangle=\frac{T_{\psi}^{\widehat{A}}(X)|\psi\rangle}{\sqrt{\langle\psi | T_{\psi}^{\widehat{A} \dag}(X)T_{\psi}^{\widehat{A}}(X) |\psi\rangle}} \ , 
\end{equation}
\emph{or, equivalently, by the density operator}
\begin{equation} \label{genpost_dis_W}
\rho_{\psi_{F}}=\frac{T_{\psi}^{\widehat{A}}(X)\rho_{\psi}T_{\psi}^{\widehat{A} \dag}(X)}{Tr[T_{\psi}^{\widehat{A}}(X)\rho_{\psi}T_{\psi}^{\widehat{A} \dag}(X)]} \ .
\end{equation}
\emph{Let the measurement yield the no outcome. Then, the state $S'_F$ of $x$ after the measurement is a pure state represented by the unit vector}
\begin{equation} \label{projpostulate_psi_no}
|\psi'_{F} \rangle=\frac{T_{\psi}^{\widehat{A}}(\Re \setminus X)|\psi\rangle}{\sqrt{\langle\psi | T_{\psi}^{\widehat{A} \dag}(\Re \setminus X)T_{\psi}^{\widehat{A}}(\Re \setminus X) |\psi\rangle}} \ , 
\end{equation}
\emph{or, equivalently, by the density operator}
\begin{equation} \label{projpostulate_W_no}
\rho_{\psi'_{F}}=\frac{T_{\psi}^{\widehat{A}}(\Re \setminus X)\rho_{\psi}T_{\psi}^{\widehat{A} \dag}(\Re \setminus X)}{Tr[T_{\psi}^{\widehat{A}}(\Re \setminus X) \rho_{\psi} T_{\psi}^{\widehat{A} \dag}(\Re \setminus X)]} \ .
\end{equation}

\vspace{.1cm}

GPP replaces the projection postulate stated in elementary textbooks and manuals on QM introducing two basic changes. Firstly, the operator $T_{\psi}^{\widehat{A}}(X)$ that depends on $|\psi\rangle$ replaces the projection operator which appears in the projection postulate and does not depend on $|\psi\rangle$. Secondly, the terms in the denominators in Eqs. (\ref{genpost_dis_psi})--(\ref{projpostulate_W_no}) do not coincide with the probabilities of the yes and no outcomes, respectively (see Eqs. (\ref{prob_X_psi}) and (\ref{prob_X_W})). 

To illustrate GPP let us consider the special case of a discrete generalized observable discussed at the end of Sect. \ref{generalizedformalism}. Whenever the property $F_n$ is measured and the yes outcome is obtained, Eq. (\ref{genpost_dis_psi}) yields
\begin{equation} \label{caso_particolare_psi_uno}
|\psi_{F_n}\rangle = \left \{
\begin{array}{cll} 
\frac{P_{n}^{\widehat{A}}|\psi\rangle}{\sqrt{\langle\psi|P_{n}^{\widehat{A}}|\psi\rangle}} & & \textrm{if} \ n \ne 0  \\
\frac{\sum_{m \in {\mathbb N}} (1-p_{\psi m}^{d}(\widehat{A}))P_{m}^{\widehat{A}}|\psi\rangle}{\sqrt{\sum_{m \in {\mathbb N}} (1-p_{\psi m}^{d}(\widehat{A}))^{2} \Vert P_{m}^{\widehat{A}}|\psi\rangle \Vert^{2}}} & & \textrm{if} \ n=0
\end{array}
\right. .
\end{equation} 
If $n \ne 0$, Eq. (\ref{caso_particolare_psi_uno}) is consistent with our assumption at the beginning of this section. If $n=0$, Eq. (\ref{caso_particolare_psi_uno}) shows that the initial state can be modified by the measurement even if the physical object is not detected, though this does not occur for special classes of generalized observables \cite{s08,sg08}.

\section{The modified BCHSH inequality\label{disgeneralizzate}}
We have already proved in some previous papers \cite{gs08,s07,g08} that, if one describes the physical situation that led to the BCHSH inequality from the point of view of the ESR model, then the conflict between the BCHSH inequality and quantum predictions disappears. This result, however, has been achieved by introducing additional assumptions and without resorting to the mathematical representation of generalized observables. We intend to show in this section that it can be restated and deepened by using the mathematical apparatus presented in the previous sections instead of introducing additional assumptions.

To begin with, let us introduce some preliminary technical remarks on joint measurements of generalized observables in the ESR model.

Let $A$ be a discrete observable of QM represented by the self--adjoint operator $\widehat{A}$, let $\{ a_1, a_2, \ldots \}$ be the set of all its possible outcomes, and let $A_0$ be a generalized observable obtained from $A$, with set of possible outcomes $\{ a_0 \} \cup \{ a_1, a_2, \ldots \}$. In this case Eq. (\ref{caso_particolare_n}) can be used to evaluate the \emph{expectation value} ${\langle A_0 \rangle}_{S}$ of $A_0$ in the pure state $S$ represented by the unit vector $|\psi\rangle$,
\begin{eqnarray}
{\langle A_0 \rangle}_{S}=\sum_{n \in {\mathbb N}_{0}}a_n p_{S}^{t}(F_n)=\sum_{n \in {\mathbb N}_{0}}a_n \langle\psi|T_{\psi}^{\widehat{A}}(\{ a_n \}) |\psi\rangle= \nonumber \\
=a_0+\sum_{n \in {\mathbb N}} (a_n-a_0) p_{\psi n}^{d}(\widehat{A})\langle\psi|P_{n}^{\widehat{A}}|\psi\rangle \ .  \label{expectation_value}
\end{eqnarray}

Let us consider another discrete observable $B$ of QM represented by the self--adjoint operator $\widehat{B}$ with set of possible outcomes $\{ b_1, b_2, \ldots \}$, let $B_0$ be a generalized observable obtained from $B$, with set of possible outcomes $\{ b_0 \} \cup \{ b_1, b_2, \ldots \}$, and let us assume that nondestructive idealized measurements of $A_0$ and $B_0$ are performed. By using GPP we can calculate the probability ${p}_{S}^{t}(a_n,b_p)$ (with $n,p \in {\mathbb N}_{0}$) of obtaining the pairs of outcomes $(a_n,b_p)$ when firstly measuring ${A}_0$ and then ${B}_0$ on a physical object $x$ in the state $S$. We get
\begin{equation} \label{assnp}
{p}_{S}^{t}(a_n,b_p)
=\langle\psi|T_{\psi}^{\widehat{A}}(\{ a_n \})|\psi\rangle \langle\psi_{F_n}|T_{\psi_{F_n}}^{\widehat{B}}(\{ b_p \})|\psi_{F_n}\rangle ,
\end{equation}
where $T_{\psi}^{\widehat{B}}(\{ b_p \})$ is given by Eq. (\ref{caso_particolare_uno}), with $p$, $b_p$ and $\widehat{B}$ in place of $n$, $a_n$ and $\widehat{A}$, respectively, and $|\psi_{F_n}\rangle$ is given by Eq. (\ref{caso_particolare_psi_uno}). Whenever $n\ne 0 \ne p$, Eq. (\ref{assnp}) yields
\begin{equation} \label{joint_prob_ne0}
{p}_{S}^{t}(a_n,b_p)=p_{\psi n}^{d}(\widehat{A})p_{\psi_{F_n} p}^{d}(\widehat{B}) \langle\psi|P_{n}^{\widehat{A}}P_{p}^{\widehat{B}}P_{n}^{\widehat{A}}|\psi\rangle \ .
\end{equation}

Let now $\Omega$ be a compound system made up of two subsystems $\Omega_1$ and $\Omega_2$, associated in standard QM with the Hilbert spaces ${\mathscr H}_{1}$ and ${\mathscr H}_{2}$, respectively, so that $\Omega$ is associated with the Hilbert space ${\mathscr H}={\mathscr H}_{1} \otimes {\mathscr H}_{2}$.

Let $A(1)$ ($B(2)$) be a discrete quantum observable of $\Omega_1$ ($\Omega_2$), with set of possible outcomes $\Xi_1=\{ a_1, a_2, \ldots \}$ ($\Xi_2=\{ b_1, b_2, \ldots \}$), represented by the self--adjoint operator $\widehat{A}(1)$ ($\widehat{B}(2)$) on ${\mathscr H}_{1}$ (${\mathscr H}_{2}$). When considered as an observable of $\Omega$, $A(1)$ ($B(2)$) is represented in standard QM by the self--adjoint operator $\widehat{A}(1) \otimes I(2)$ ($I(1) \otimes \widehat{B}(2)$), where $I(2)$ ($I(1)$) is the identity operator on ${\mathscr H}_{2}$ (${\mathscr H}_{1}$), that we still denote by $\widehat{A}(1)$ ($\widehat{B}(2)$) for the sake of simplicity. Let $A_0(1)$ ($B_0(2)$) be a generalized observable obtained from $A(1)$ ($B(2)$) by adding the no--registration outcome $a_0$ ($b_0$) to $\Xi_1$ ($\Xi_2$). Whenever simultaneous measurements of $A_0(1)$ and $B_0(2)$ are performed on a physical object $x$ (individual example of $\Omega$) in a pure state $S$ such that $\Omega_1$ and $\Omega_2$ are spatially separated, noncontextuality implies that the transformation of $S$ induced by the measurement of $A_0(1)$ must not affect the detection probability associated with the measurement of $B_0(2)$. If $S$ is represented by the unit vector $|\Psi\rangle \in {\mathscr H}$, we obtain
\begin{equation}
{p}_{\Psi_{F_n} p}^{d}(\widehat{B}(2))={p}_{\Psi p}^{d}(\widehat{B}(2)) ,
\end{equation}
hence  Eq. (\ref{joint_prob_ne0}) yields  
\begin{equation} \label{assnp1}
{p}_{S}^{t}(a_n,b_p)=p_{\Psi n}^{d}(\widehat{A}(1))p_{\Psi p}^{d}(\widehat{B}(2)) \langle\Psi|P_{n}^{\widehat{A}(1)} P_{p}^{\widehat{B}(2)}|\Psi\rangle \ .
\end{equation}

We can now define the expectation value of the product of the generalized observables $A_0(1)$ and $B_0(2)$ in the state $S$ as follows,
\begin{eqnarray} 
E(A_0(1),B_0(2))=\sum_{n,p \in {\mathbb N}} a_n b_p {p}_{S}^{t}(a_n,b_p)+ \nonumber \\
+\sum_{n\in {\mathbb N}} a_n b_0 {p}_{S}^{t}(a_n,b_0)+ \sum_{p\in {\mathbb N}} a_0 b_p {p}_{S}^{t}(a_0,b_p)+a_0 b_0 {p}_{S}^{t}(a_0,b_0). \label{corrfunc}  
\end{eqnarray}
By using Eq. (\ref{assnp1}) and restricting to generalized observables such that $a_0=0=b_0$\footnote{Note that, for every generalized observable $A_0$, with $a_0 \ne 0$, one can construct a new observable whose no--registration outcome is $0$. Indeed, one can select a Borel function on $\Re$ which is bijective on $\Xi_0$ and such that $\chi(a_0)=0$, and consider the generalized observable $\chi(A_0)$ obtained from $\chi(A)$ by adjoining the outcome $0$ and putting, for every $\lambda \in \Re$, ${p}_{\psi}^{d}(\chi(\widehat{A}), \lambda)={p}_{\psi}^{d}(\widehat{A}, \chi^{-1}(\lambda))$ (hence $p_{S}^{t}(\chi(A_0), \{0 \})=p_{S}^{t}(A_0, \{ a_0 \})$ because of Eq. (\ref{normalization_lebesgue})).\label{no_registration_0}} (hence, for every $n,p \in {\mathbb N}$, $a_n \ne 0 \ne b_p$) we get
\begin{equation}
E(A_0(1),B_0(2))
=\sum_{n,p \in {\mathbb N}} a_n b_p p_{\Psi n}^{d}(\widehat{A}(1))p_{\Psi p}^{d}(\widehat{B}(2)) \langle\Psi|P_{n}^{\widehat{A}(1)} P_{p}^{\widehat{B}(2)}|\Psi\rangle \ ,  \label{corrfunc_ass}
\end{equation}
because, obviously, $P_{n}^{\widehat{A}(1)}$ and $P_{p}^{\widehat{B}(2)}$ commute.

Let us come to our main aim in this section and recall the notion of \emph{local realism} as usually understood in the literature \cite{b64,chsh69,epr35}. To be precise, this notion indicates the join of the assumptions of ``realism'' (\emph{R}) and ``locality'' (\emph{LOC}):

\emph{R}: \emph{the values of all observables of a physical system in a given state are predetermined for any measurement context};

\emph{LOC}: \emph{if measurements are made at places remote from one another on parts of a physical system which no longer interact, the specific features of one of the measurements do not influence the results obtained with the others}.\footnote{The use of the term \emph{local realism} has been recently criticized by Norsen \cite{n07}, mainly because R does not comply with any definition of realism in the philosophical literature. Nevertheless we think that it is clearly defined by R and LOC and that it can be maintained as a conventional locution because its use is widespread in physics.}

Then the standard procedures leading to the BCHSH inequality can be resumed as follows. One considers a compound physical system $\Omega$ made up of two far away subsystems $\Omega_1$ and $\Omega_2$, and two dichotomic observables  $A({\bf a})$ and $B({\bf b})$ of $\Omega_1$ and $\Omega_2$, respectively, depending on the parameters ${\bf a}$ and ${\bf b}$ and taking either value $-1$ or $1$. The expectation value $E({\bf a}, {\bf b})$ of the product of the observables $A({\bf a})$ and $B({\bf b})$ in a state $S$ is given by
\begin{equation} \label{lhvt}
E({\bf a}, {\bf b})= \int_{\Lambda} d\lambda \rho(\lambda) A(\lambda, {\bf a}) B(\lambda, {\bf b}),
\end{equation}
where $\lambda$ is a deterministic \emph{hidden variable} whose value ranges over a domain $\Lambda$ when measurements on different examples of $\Omega$ in the state $S$ are considered, $\rho(\lambda)$ is a probability distribution on $\Lambda$, $A(\lambda, {\bf a})$ and $B(\lambda, {\bf b})$ are the values of $A({\bf a})$ and $B({\bf b})$, respectively. By assuming R and LOC one easily gets
\begin{equation} \label{chsh_69}
|E({\bf a}, {\bf b})-E( {\bf a}, {\bf b'})|+|E({\bf a'}, {\bf b})+E({\bf a'}, {\bf b'})| \le 2.
\end{equation}
It is now important to note that the proof of Eq. (\ref{chsh_69}) requires the assumption, usually left implicit, that ideal measurements are performed in which all physical objects that are prepared are also detected.\footnote{Actual measurements usually do not fulfill this condition, hence the BCHSH inequality cannot be tested directly. Empirical tests refer to derived inequalities, obtained from the BCHSH inequality by adding some further assumptions to R and LOC. The reliability of these assumptions is disputed by many authors, who therefore uphold that the empirical data that show that the derived inequalities are violated do not prove that R and LOC do not hold \cite{sf02,f82,f89,s04,s05,dcg96,s00,gg99}. We refer to \cite{gs08,g07} for a more detailed analysis of this topic and comparison with the perspective introduced by the ESR model.} Indeed this condition does not hold in the ESR model, where the dichotomic observables $A({\bf a})$, $B({\bf b})$, $A({\bf a'})$, $B({\bf b'})$ must be substituted by the trichotomic generalized observables $A_0({\bf a})$, $B_0({\bf b})$, $A_0({\bf a'})$, $B_0({\bf b'})$, respectively, in each of which a no--registration outcome is adjoined to the outcomes $+1$ and $-1$. Hence, the reasonings that lead to Eq. (\ref{chsh_69}) must be modified if the perspective introduced by the ESR model is adopted. We have discussed this issue in a recent paper, by using the microscopic part of the ESR model that provides a hidden variables theory for QM (with reinterpretation of quantum probabilities) and assuming that all no--registration outcomes are $0$ (see footnote \ref{no_registration_0}). By denoting the expectation value of the product of $A_0({\bf a})$ and $B_0({\bf b})$ in the state $S$ (see Eq. (\ref{corrfunc})) by $E(A_0({\bf a}), B_0({\bf b}))$ we have shown that the following \emph{modified BCHSH inequality} holds \cite{gs08,s07,g08}
\begin{equation} \label{modified_bchsh} 
|E(A_0({\bf a}), B_0({\bf b}))-E( A_0({\bf a}), B_0({\bf b'}))|
+|E(A_0({\bf a'}), B_0({\bf b}))+E(A_0({\bf a'}), B_0({\bf b'}))| \le 2 , 
\end{equation}
which replaces Eq. (\ref{chsh_69}) within the ESR model.

Coming to our present framework, one can particularize Eq. (\ref{corrfunc_ass}) to trichotomic generalized observables that can take only values $+1$, $0$ and $-1$, and substitute it into Eq. (\ref{modified_bchsh}). The resulting equation, however, is still too general and complicate for our present purposes. Therefore, let us assume that the set ${\mathcal O}_{R}$ of generalized observables such that, for every $A_0 \in {\mathcal O}_{R}$, the detection probability in a given state depends on $A_0$ but not on its specific value is non--void, and let us restrict to ${\mathcal O}_{R}$. Hence we can drop the dependence on $n$ and $p$ of the detection probabilities that appear in Eq. (\ref{corrfunc_ass}) and get from Eq. (\ref{corrfunc_ass})
\begin{eqnarray}
E(A_0({\bf a}),B_0({\bf b}))=p_{\Psi}^{d}(\widehat{A}({\bf a})) p_{\Psi}^{d}(\widehat{B}({\bf b})) [\langle\Psi|P_{1}^{\widehat{A}({\bf a})} P_{1}^{\widehat{B}({\bf b})}|\Psi\rangle+ \nonumber \\
-\langle\Psi|P_{1}^{\widehat{A}({\bf a})} P_{-1}^{\widehat{B}({\bf b})}|\Psi\rangle
-\langle\Psi|P_{-1}^{\widehat{A}({\bf a})} P_{1}^{\widehat{B}({\bf b})}|\Psi\rangle+ \nonumber \\
+\langle\Psi|P_{-1}^{\widehat{A}({\bf a})} P_{-1}^{\widehat{B}({\bf b})}|\Psi\rangle]= p_{\Psi}^{d}(\widehat{A} ({\bf a})) p_{\Psi}^{d}(\widehat{B}({\bf b}))\langle \widehat{A}({\bf a})\widehat{B}({\bf b})\rangle_{\Psi} \label{expectation_value_spin}
\end{eqnarray}
where $\langle \widehat{A}({\bf a})\widehat{B}({\bf b})\rangle_{\Psi}$ is the quantum expectation value, in the state $S$, of the product of the quantum observables $A({\bf a})$ and $B({\bf b})$ from which $A_0({\bf a})$ and $B_0({\bf b})$, respectively, are obtained.\footnote{We stress that $\langle\widehat{A}({\bf a})\widehat{B}({\bf b})\rangle_{\psi}$ is interpreted as a conditional expectation value in the ESR model, that is, as the mean value of the product of the generalized observables $A_0({\bf a})$ and $B_0({\bf b})$ whenever only detected objects are taken into account.} Since similar equations hold if we consider $A_0({\bf a})$ and $B_0({\bf b'})$, $A_0({\bf a'})$ and $B_0({\bf b})$, $A_0({\bf a'})$ and $B_0({\bf b'})$, we obtain from Eq. (\ref{modified_bchsh})
\begin{eqnarray}
p_{\Psi}^{d}(\widehat{A}({\bf a})) |p_{\Psi}^{d}(\widehat{B}({\bf b}))\langle \widehat{A}({\bf a})\widehat{B}({\bf b})\rangle_{\Psi}
- p_{\Psi}^{d}(\widehat{B}({\bf b'}))\langle \widehat{A}({\bf a})\widehat{B}({\bf b'})\rangle_{\Psi}|+ \nonumber \\
+p_{\Psi}^{d}(\widehat{A}({\bf a'})) | p_{\Psi}^{d}(\widehat{B}({\bf b}))\langle \widehat{A}({\bf a'})\widehat{B}({\bf b})\rangle_{\Psi}
+ p_{\Psi}^{d}(\widehat{B}({\bf b'}))\langle \widehat{A}({\bf a'})\widehat{B}({\bf b'})\rangle_{\Psi}|\le 2 . \label{modified_bchsh_assumption}
\end{eqnarray} 
Eq. (\ref{modified_bchsh_assumption}) constitutes our main result in this section and deserves some comments.\footnote{The result expressed by Eq. (\ref{modified_bchsh_assumption}) can be obtained by introducing additional assumptions, without using the mathematical representation of generalized observables introduced by the ESR model and GPP \cite{gs08,s07}. Our present treatment avoids such assumptions and recovers Eq. (\ref{modified_bchsh_assumption}) in the general mathematical framework of the ESR model.}

First of all, we note that the procedure leading to Eq. (\ref{modified_bchsh_assumption}) corresponds in the ESR model to the standard procedure of substituting quantum expectation values in the BCHSH inequality. It is well known that this substitution leads to contradictions if states and observables are suitably chosen. In the ESR model, instead, four detection probabilities appear, whose value is \emph{a priori} unknown. From a logical point of view one cannot conclude that local realism is not compatible with QM. Rather, Eq. (\ref{modified_bchsh_assumption}) must be interpreted as a condition that has to be fulfilled by the detection probabilities in the ESR model, which makes the ESR model falsifiable, at least in principle. Indeed, we have no theory, at present, which allows us to predict the values of the detection probabilities, but if one can perform measurements that are close to ideality the detection probabilities can be determined experimentally. If the obtained values are such that the inequality in Eq. (\ref{modified_bchsh_assumption}) is fulfilled, the ESR model is confirmed, and no contradiction occurs between local realism and the mathematical apparatus of QM. If the obtained values are such that the inequality is violated, the ESR model, or the simplificative assumption that ${\mathcal O}_{R}$ is non--void, or both, are falsified.

Secondly, we recall that Eq. (\ref{modified_bchsh}), hence Eq. (\ref{modified_bchsh_assumption}), has been obtained by referring to the set of all physical objects that are prepared. If one refers instead to the set of all objects that are detected, quantum predictions hold because of assumption AX (Sect. \ref{modello}) which are not consistent with the BCHSH inequality. Thus, the BCHSH inequality does not hold in both cases that can be experimentally investigated according to the ESR model. But this occurs in a framework in which local realism holds, which contradicts orthodox beliefs.

Thirdly, we note that, if one considers the microscopic part of the ESR model and introduces microscopic, purely theoretical, observables, the BCHSH inequality holds for such observables \cite{gs08,g08}. Thus one obtains ``conciliatory'' results in the ESR model, in the sense that the BCHSH inequality, the modified BCHSH inequality and the quantum inequalities do not conflict, but rather pertain to different parts of the picture provided by the ESR model.

The above remarks can be illustrated by dealing with special cases. In particular, we have recently shown that, under reasonable physical assumptions, spin measurements on a system of two far apart spin--$\frac{1}{2}$ quantum particles in the singlet spin state 
cannot have a detection efficiency greater than $0.841$ \cite{s07}. Should this statement be contradicted by experimental data, the ESR model, or the ``reasonable'' assumptions that have been introduced, or both, would be falsified. If not, one can consider this result as a clue that the ESR model is correct.

\section{Generalized observables and mixtures\label{gen_obse_mixed_case}}
Assumption AX in Sect. \ref{modello} allows one to calculate the probability $p_{S}(F)$ in Eq. (\ref{formuladipartenza}) by using standard QM rules whenever $S$ is a pure state. However, Eq. (\ref{formuladipartenza}) has been derived without making assumptions on $S$, hence it holds also if $S$ is a mixture. One is thus led to wonder whether $p_{S}(F)$ can be calculated by means of standard quantum rules also in this case. We intend to show in the present section that the answer is negative and to provide new rules for evaluating $p_{S}(F)$ and $p_{S}^{t}(F)$ in the case of mixtures by using the mathematical representations introduced in Sect. \ref{generalizedformalism}.

Let $S$ be a mixture of the pure states $S_1, S_2, \ldots$, represented by the density operators $\rho_{\psi_1}$, $\rho_{\psi_2}$, \ldots, with probabilities $p_1, p_2, \ldots$, respectively. The probability that a measurement of the generalized observable $A_0$ on a physical object $x$ in the state $S$ yield an outcome in the Borel set $X \in \mathbb{B}(\Re)$, with $a_0 \notin X$, coincides with the probability $p_{S}^{t}(F)$ that $x$ display the property $F=(A_0, X) \in \mathcal F$ in the state $S$. Because of Assumption AX we get
\begin{equation} \label{prob_tot_mix_phys}
p_{S}^{t}(F)=\sum_{j} p_{j} p_{S_j}^{t}(F)=\sum_{j} p_j p_{S_j}^{d}(F) p_{S_j}(F),
\end{equation} 
where $p_{S_j}^{t}(F)$ is the overall probability that a physical object $x$ in the pure state $S_{j}$ display $F$ when an (idealized) measurement of $F$ is performed on $x$, $p_{S_j}^{d}(F)$ is the probability that $x$ be detected and $p_{S_j}(F)$ is the conditional probability that $x$ display $F$ when detected. Because of Eq. (\ref{formuladipartenza}) we get
\begin{equation} \label{prob_mix_phys}
p_{S}(F)=\sum_{j} p_{j} \frac{p_{S_j}^{d}(F)}{p_{S}^{d}(F)} p_{S_j}(F).
\end{equation}
Eq. (\ref{prob_mix_phys}) is reasonable from an intuitive point of view. Indeed, bearing in mind the interpretation of the probabilities that appear in it, the term $p_j \frac{p_{S_j}^{d}(F)}{p_{S}^{d}(F)}$ can be interpreted, because of the Bayes theorem, as the conditional probability that $x$ be in the state $S_j$ whenever $F$ is measured and $x$ is detected. 

Eq. (\ref{prob_mix_phys}) can be rewritten by using the mathematical representations of generalized observables provided in Sect. \ref{generalizedformalism}. Indeed, assumption AX yields 
\begin{equation} \label{prob_pure_math}
p_{S_j}(F)=Tr[\rho_{\psi_j} P^{\widehat{A}}(X)],
\end{equation}
where $P^{\widehat{A}}$ is the (spectral) PV measure associated with the self--adjoint operator $\widehat{A}$ representing the observable $A$ of QM from which $A_0$ is obtained. Hence
\begin{equation} \label{AX_mixed_0}
p_{S}(F)=Tr \Big [(\sum_{j} p_j \frac{p_{S_j}^{d}(F)}{p_{S}^{d}(F)} \rho_{\psi_j}) P^{\widehat{A}}(X) \Big ] \ .
\end{equation}
Furthermore, if one introduces the obvious assumption
\begin{equation} \label{prob_mix_d}
p_{S}^{d}(F)=\sum_{j} p_j p_{S_j}^{d}(F)
\end{equation}
and uses Eq. (\ref{prova_trattazione_nuova_rho}) (which holds for pure states only) with $S_j$ in place of $S$, one gets
\begin{equation}
\frac{p_{S_j}^{d}(F)}{p_{S}^{d}(F)}=\frac{
\frac{Tr[\rho_{\psi_j}T_{\psi_j}^{\widehat{A}}(X)]}{Tr[\rho_{\psi_j} P^{\widehat{A}}(X)]}}{\sum_{j} p_j \frac{Tr[\rho_{\psi_j}T_{\psi_j}^{\widehat{A}}(X)]}{Tr[\rho_{\psi_j} P^{\widehat{A}}(X)]}} \ ,
\end{equation}
hence
\begin{equation} \label{AX_mixed_1}
p_{S}(F)=Tr[\rho_{S}(F)P^{\widehat{A}}(X) ] \ , 
\end{equation}
with
\begin{equation} \label{rho_S(F)}
\rho_{S}(F)=\sum_{j} p_j \frac{p_{S_j}^{d}(F)}{p_{S}^{d}(F)} \rho_{\psi_j}=  \frac{\sum_{j} p_j
\frac{Tr[\rho_{\psi_j}T_{\psi_j}^{\widehat{A}}(X)]}{Tr[\rho_{\psi_j} P^{\widehat{A}}(X)]} \rho_{\psi_j}}{\sum_{j} p_j \frac{Tr[\rho_{\psi_j}T_{\psi_j}^{\widehat{A}}(X)]}{Tr[\rho_{\psi_j} P^{\widehat{A}}(X)]}}  \ .
\end{equation}
Eqs. (\ref{AX_mixed_1}) and (\ref{rho_S(F)}) show that $p_{S}(F)$ does not coincide, in general, with the probability obtained by applying standard QM rules, that is, calculating $Tr[\rho_{S} P^{\widehat{A}}(X)]$, with $\rho_{S}=\sum_{j} p_j \rho_{\psi_{j}}$. Intuitively, this can be explained by observing that, whenever a physical property $F$ is measured on an ensemble of physical objects prepared in the state $S$, the ensemble of detected objects depends on $F$ and generally is not a fair sample of the set of all prepared objects. Hence, as far as $p_{S}(F)$ is concerned, $S$ must be represented by the density operator $\rho_{S}(F)$, which depends on $F$ and coincides with $\rho_{S}$ only in special cases. More generally, $S$ must be associated with the family of density operators 
\begin{equation}
\{\rho_{S}(F)  \}_{F \in {\mathcal F}}= \{ \sum_{j} p_j \frac{p_{S_j}^{d}(F)}{p_{S}^{d}(F)} \rho_{\psi_j} \}_{F \in {\mathcal F}} ,
\end{equation} 
which provides a representation of $S$ in the ESR model. If pure states are considered as limiting cases of mixtures, this family reduces to the constant $\{\rho_{S} \}_{F \in {\mathcal F}}$ whenever $S$ is a pure state, which implies that Eq. (\ref{AX_mixed_1}) embodies assumption AX. But Eq. (\ref{AX_mixed_1}) also shows that assumption AX cannot be extended to nonpure states. 

Let us come to the probability $p_{S}^{t}(F)$. By using Eq. (\ref{prob_X_W}) we get
\begin{equation}
p_{S_j}^{t}(F)=Tr[\rho_{\psi_j}T_{\psi_j}^{\widehat{A}}(X)],
\end{equation} 
where $T_{\psi_j}^{\widehat{A}}(X)=\int_{X}{p}_{\psi_j}^{d}(\widehat{A}, \lambda) \mathrm{d} P^{\widehat{A}}_{\lambda}$ because $a_0 \notin X$. Hence
\begin{equation} \label{prob_tot_mix_math_notin}
p_{S}^{t}(F)=Tr \Big [\sum_{j} p_j \rho_{\psi_j} T_{\psi_j}^{\widehat{A}}(X) \Big ].
\end{equation}
Eq. (\ref{prob_tot_mix_math_notin}) shows that, generally, $p_{S}^{t}(F)$ cannot be written as the trace of the product of two operators, one of which represents $S$ and the other represents $F$. However, by using Eqs. (\ref{prova_trattazione_nuova_rho}) and (\ref{rho_S(F)}) we get from Eq. (\ref{prob_tot_mix_math_notin})
\begin{equation} \label{conferma}
p_{S}^{t}(F)=Tr \Big [\sum_j p_j \rho_{\psi_j} p_{S_j}^{d}(F)  P^{\widehat{A}}(X) \Big ]
=p_{S}^{d}(F) Tr \Big [\rho_{S}(F) P^{\widehat{A}}(X) \Big ] \ ,
\end{equation}
consistently with Eq. (\ref{formuladipartenza}).

Let us now consider the property $G=(A_0, Y) \in {\mathcal F}_{0} \setminus \mathcal F$, hence $a_0 \in Y$, and put $F=(A_0, X)$, with $X= Y \setminus \{ a_0 \}$. We get from Eqs. (\ref{f_bar_c}), (\ref{prob_tot_mix_phys}) and (\ref{prob_mix_d}) 
\begin{eqnarray}
p_{S}^{t}(G)=1-p_{S}^{d}(F)+p_{S}^{t}(F)
=\sum_{j}p_j-\sum_{j} p_j p_{S_j}^{d}(F)+\sum_{j} p_j p_{S_j}^{t}(F)= \nonumber \\
=\sum_{j}p_j(1-p_{S_j}^{d}(F)+p_{S_j}^{t}(F))=\sum_{j} p_j p_{S_j}^{t}(G).
\end{eqnarray}
By using Eq. (\ref{prob_X_W}) we obtain
\begin{equation}
p_{S_j}^{t}(G)=Tr[\rho_{\psi_{j}} T_{\psi_j}^{\widehat{A}}(Y)]
\end{equation}
where $T_{\psi_j}^{\widehat{A}}(Y)=I - \int_{\Re \setminus Y}{p}_{\psi_j}^{d}(\widehat{A}, \lambda)\mathrm{d} P^{\widehat{A}}_{\lambda}$ because of Eq. (\ref{POV_lebesgue}), hence
\begin{equation} \label{prob_tot_mix_math_in}
p_{S}^{t}(G)=Tr \Big [ \sum_{j} p_{j} \rho_{\psi_j} T_{\psi_j}^{\widehat{A}}(Y) \Big ].
\end{equation}
Putting together Eqs. (\ref{prob_tot_mix_math_notin}) and (\ref{prob_tot_mix_math_in}), we finally obtain that, for every property $(A_0, X)$, with $X \in {\mathbb B}(\Re)$,
\begin{equation} \label{prob_tot_mix_math_gen}
p_{S}^{t}((A_0, X))=Tr \Big [\sum_{j} p_{j} \rho_{\psi_j} T_{\psi_j}^{\widehat{A}}(X) \Big ],
\end{equation}
where $T_{\psi_j}^{\widehat{A}}(X)$ is given by Eq. (\ref{POV_lebesgue}) with $\psi_j$ in place of $\psi$.

\section{The ignorance interpretation of mixtures\label{ignorance}}
Let $S$ be the mixture introduced in Sec. \ref{gen_obse_mixed_case}, which is represented in QM by the density operator $\rho_S=\sum_{j} p_j \rho_{\psi_j}$. Then, a typical preparation procedure a physical object $x$ in the state $S$ can be summarized as follows.

\emph{Select a preparing device $\pi_{j}$ for every pure state $S_{j}$, use $\pi_j$ to prepare an ensemble ${\mathscr E}_{S_j}$ of $n_j$ physical objects in the state $S_j$, mingle the ensembles ${\mathscr E}_{S_1}, {\mathscr E}_{S_2}, \ldots$ to prepare an ensemble ${\mathscr E}_{S}$ of $N=\sum_{j} n_j$ physical objects and assume that each $n_j$ is such that $n_j=N p_{j}$. Then, remove any memory of the way in which the ensembles ${\mathscr E}_{S_1}, {\mathscr E}_{S_2}, \ldots$ have been mingled and select a physical object $x$ in ${\mathscr E}_{S}$}.

The class of preparation procedures obtained proceeding as above and selecting the preparing devices in the states $S_1$, $S_2$, \ldots in all possible ways will be denoted by $\sigma_S$ and called \emph{operational definition} of $S$ in the following.

The operational definition of $S$ implies that the probability $p_j$ is \emph{epistemic}, that is, it can be interpreted as formalizing the loss of memory about the pure state in which each physical object has been actually prepared (\emph{ignorance interpretation} of $p_j$). It is well known, however, that there generally exist one--dimensional projection operators $\rho_{\chi_1}, \rho_{\chi_2}, \ldots$, none of which coincides with one of the projection operators $\rho_{\psi_1}, \rho_{\psi_2}, \ldots$, which are such that $\rho_S=\sum_{l} q_l \rho_{\chi_l}$, with $0\le q_l \le 1$ and $\sum_{l} q_l=1$. If this expression of $\rho_S$ is adopted, the coefficients $q_l$ cannot be interpreted as probabilities bearing an ignorance interpretation. 

The nonunique decomposition of quantum mixtures is usually considered a distinguishing feature of QM, but it is a source of interpretative problems. In particular, consider a mixture $T$ of the pure states $T_1, T_2, \ldots$ represented by the density operators $\rho_{\chi_1}, \rho_{\chi_2}, \ldots$, with probabilities $q_1, q_2, \ldots$, respectively, prepared following the procedure described in the case of $S$ with obvious changes, so that $T$ has an operational definition $\sigma_T$ which is different from $\sigma_S$. According to QM the state $T$ is physically equivalent to $S$ because it is represented by the same density operator, hence it must be identified with $S$. But the probabilities $q_1, q_2, \ldots$ now admit an ignorance interpretation, at variance with the conclusion expounded above. Many scholars therefore maintain that an ignorance interpretation of the probabilities that appear in the various possible expressions of $\rho_S$ must be avoided \cite{bc81}, which however clashes with the interpretation of these probabilities in the operational definitions of $S$ and $T$. More rigorously, the definition of state in QM identifies two mixtures $S$ and $T$ if and only if they are \emph{probabilistically equivalent} (\emph{i.e.}, they associate the same probability with every property of the physical system that is considered), which occurs, because of the Gleason theorem, if and only if $\rho_S=\rho_T$. But, then, the representation of $S$ and $T$ by means of density operators cannot distinguish between the probabilistically equivalent but pragmatically different operational definitions $\sigma_S$ and $\sigma_T$ \cite{blm91}. 

Let us come to the ESR model. As in QM, the definition of state implies that two mixtures $S$ and $T$ whose operational definitions are different must be identified if and only if $\sigma_{S}$ and $\sigma_T$ are probabilistically equivalent ($\sigma_S \equiv \sigma_T$), that is, if and only if, for every $F \in {\mathcal F}_{0}$, $p_{S}^{t}(F)=p_{T}^{t}(F)$. It follows from Eqs. (\ref{formuladipartenza}) and (\ref{f_bar_c}) that $\sigma_S \equiv \sigma_T$ if and only if, for every $F \in {\mathcal F}$, $p_{S}^{d}(F)=p_{T}^{d}(F)$ and $p_{S}(F)=p_{T}(F)$. Because of Eq. (\ref{AX_mixed_1}), the latter condition holds if and only if, for every $F \in {\mathcal F}$, the equality $Tr[\rho_{S}(F)P^{\widehat{A}}(X) ]=Tr[\rho_{T}(F)P^{\widehat{A}}(X) ]$ holds. This equality does not imply that the family of density operators associated with $S$ coincides with the family associated with $T$. By construction, indeed, $\{\rho_{S}(F) \}_{F \in {\mathcal F}}$ and $ \{\rho_{T}(F) \}_{F \in {\mathcal F}}$ are determined by the operational definitions $\sigma_{S}$ and $\sigma_T$, respectively. Hence it may occur that $\{\rho_{S}(F) \}_{F \in {\mathcal F}} \ne \{\rho_{T}(F) \}_{F \in {\mathcal F}}$ even if $\sigma_S \equiv \sigma_T$. Let us introduce now in the ESR model the following assumptions.

(i) $\sigma_{S} \ne \sigma_{T}$ $\Longrightarrow$ $\{\rho_{S}(F) \}_{F \in {\mathcal F}} \ne \{\rho_{T}(F) \}_{F \in {\mathcal F}}$;

(ii) $\sigma_{S} \equiv \sigma_{T}$ $\Longrightarrow$ $\rho_S = \rho_T$.

\noindent
Assumptions (i) and (ii) require, intuitively, that the detection probabilities $p_{S_1}^{d}(F)$, $p_{S_2}^{d}(F)$, \ldots take a sufficiently large number of values when $F$ varies in $\mathcal F$, which is physically reasonable (note that assumption (ii) is not \emph{a priori} true, because in the ESR model $\sigma_S \equiv \sigma_T$ does not imply that $Tr[\rho_{S}P^{\widehat{A}}(X) ]=Tr[\rho_{T}P^{\widehat{A}}(X) ]$ for every $F \in {\mathcal F}$; also note that we do not assume in (ii) that the converse implication holds). Then, we can draw the commutative diagram 
\begin{equation} \label{diagram}
\centerline{\includegraphics[scale=0.7]{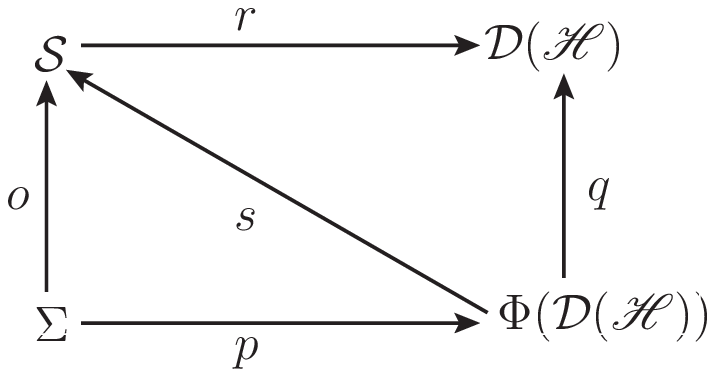}}
\end{equation}

\vspace{.1cm}
\noindent
The symbols in diagram (\ref{diagram}) are defined as follows.

$\Sigma$ is the set of all operational definitions of states. 

$\mathcal {D}({\mathscr H})$ is the convex set of all density operators on ${\mathscr H}$.

 $\Phi(\mathcal {D}({\mathscr H}))$ is the set of all families of the form $\{\rho_{S}(F) \}_{F \in {\mathcal F}}$, with $S \in {\mathcal S}$.

$o: \sigma_S \in \Sigma \longmapsto S \in {\mathcal S}$ maps each class of probabilistically equivalent operational definitions into a state. 

$p: \sigma_S \in \Sigma \longmapsto \{\rho_{S}(F)\}_{F \in {\mathcal F}} \in \Phi(\mathcal {D}({\mathscr H}))$ is defined via Eq. (\ref{rho_S(F)}). 

$r:S \in {\mathcal S} \longmapsto \rho_{S} \in \mathcal {D}({\mathscr H})$ is the \emph{standard representation} of $\mathcal S$. 

$s: \{\rho_{S}(F)\}_{F \in {\mathcal F}} \in \Phi(\mathcal {D}({\mathscr H})) \longmapsto  S \in {\mathcal S}$ is the mapping that makes the lower triangle commutative.

$q: \{\rho_{S}(F)\}_{F \in {\mathcal F}} \in \Phi(\mathcal {D}({\mathscr H})) \longmapsto \rho_{S} \in \mathcal {D}({\mathscr H})$ is the mapping that makes the upper triangle commutative. 

The mapping $p$ is bijective because of assumption (i). All remaining mappings are generally non--bijective. 

It is apparent that diagram (\ref{diagram}) offers a solution of the interpretative problem discussed above. Indeed, $q^{-1}(\rho_S)$ is a set of families that correspond, via $p$, to operational definitions (which need not be probabilistically equivalent because $\rho_S = \rho_T$ does not generally imply $S \equiv T$ in the ESR model). Specifying $\rho_S$ is then insufficient to single out an operational definition, which generates the interpretative ambiguities occurring in QM. These ambiguities do not occur in the ESR model, where the representation of a state by means of a family of density operators makes reference to a specific operational definition and not only to an equivalence class of operational definitions.\footnote{We stress that diagram (\ref{diagram}) shows that the ESR model introduces new mathematical objects representing operational definitions, hence, for every state $S$, subclasses of preparation procedures in the equivalence class of all preparation procedures defining $S$. This suggests that the standard equivalence relations based on probability should be refined to take into account some physically relevant operational differences (see also \cite{gs07b}).}

\section{State transformations induced by idealized nondestructive measurements\label{gpp_mixed_case}}
We have seen in Sect. \ref{genprojpost_measproc} that GPP rules the transformation of a pure state induced by an idealized nondestructive measurement. We now intend to show that our results in Sect. \ref{gen_obse_mixed_case}, together with GPP, allow us to predict the transformation of a mixture induced by a measurement of the same kind.

Let $S$ be the mixture introduced in Sect. \ref{gen_obse_mixed_case}, whose standard representation is provided by the density operator $\rho_{S}=\sum_{j} p_j \rho_{\psi_j}$.  Let $F=(A_0, X) \in {\mathcal F}_{0}$ be any property of $\Omega$. Whenever an idealized measurement of $F$ is performed on a physical object $x$ in the state $S$, the probabilities $p_{S_j}^{t}(F)$ and $p_{S}^{t}(F)$ can be deduced from Eq. (\ref{prob_tot_mix_math_gen}). If the measurement yields the yes outcome, the final state $S_F$ is a mixture of the pure states $S_{1F}, S_{2F}, \ldots$ represented by the density operators $\rho_{\psi_{1F}}, \rho_{\psi_{2F}}, \ldots$, respectively, obtained by applying Eq. (\ref{genpost_dis_W}), with probabilities $p_{1F}$, $p_{2F}$, \ldots, respectively, obtained by using the Bayes theorem (indeed, $p_{jF}$, with $j=1,2, \ldots$ denotes the conditional probability that $x$ be in the state $S_{jF}$ whenever a measurement of $F$ on $x$ has yielded the yes outcome). Hence GPP can be extended to mixtures, taking the form of a \emph{generalized L\"{u}ders postulate}, as follows.

\vspace{.1cm}
\noindent
\emph{GLP}. \emph{Let $S$ be a mixture of the pure states $S_1, S_2, \ldots$, represented by the density operators $\rho_{\psi_1}$, $\rho_{\psi_2}$, \ldots, with probabilities $p_1, p_2, \ldots$, respectively, and let a nondestructive idealized measurement of the physical property $F=(A_0,X)\in {\mathcal F}_{0}$ be performed on a physical object $x$ in the state $S$}.

\emph{Let the measurement yield the yes outcome. Then, the state $S_F$ of $x$ after the measurement is a mixture of the pure states $S_{1F}, S_{2F}, \ldots$ represented by the density operators $\rho_{\psi_{1F}}, \rho_{\psi_{2F}}, \ldots$, respectively, with}
\begin{equation} \label{GLP'_yes}
\rho_{\psi_{j F}}=\frac{T_{\psi_j}^{\widehat{A}}(X)\rho_{\psi_j}T_{\psi_{j}}^{\widehat{A} \dag}(X)}{Tr[T_{\psi_j}^{\widehat{A}}(X) \rho_{\psi_j} T_{\psi_j}^{\widehat{A} \dag}(X)]} \ ,
\end{equation}
\emph{and probabilities $p_{1F}$, $p_{2F}$, \ldots, respectively, with}
\begin{equation} \label{prob_bayes_yes}
p_{j F}=p_j \frac{p_{S_j}^{t}((A_0, X))}{p_{S}^{t}((A_0, X))}= p_{j} \frac{Tr[\rho_{\psi_j} T_{\psi_j}^{\widehat{A}}(X)]}{Tr \Big [\sum_{j} p_{j} \rho_{\psi_j} T_{\psi_j}^{\widehat{A}}(X) \Big ]} \ ,
\end{equation}
\emph{hence $S_F$ is represented by the family of density operators} 
\begin{equation} \label{family_GLP_yes}
\{ \rho_{S_F}(H) \}_{H \in {\mathcal F}}= \{\sum_{j} p_{j F} \frac{p_{S_j F}^{d}(H)}{p_{S_F}^{d}(H)} \rho_{\psi_{j F}} \}_{H \in {\mathcal F}} \ .
\end{equation}

\emph{Let the measurement yield the no outcome. Then, the state $S'_{F}$ of $x$ after the measurement is a mixture of the pure states $S'_{1F}, S'_{2F}, \ldots$ represented by the density operators $\rho_{\psi'_{1F}}, \rho_{\psi'_{2F}}, \ldots$, respectively, with}
\begin{equation} \label{GLP'_no}
\rho_{\psi'_{j F}}=\frac{T_{\psi_j}^{\widehat{A}}(\Re \setminus X)\rho_{\psi_j}T_{\psi_{j}}^{\widehat{A} \dag}(\Re \setminus X)}{Tr[T_{\psi_j}^{\widehat{A}}(\Re \setminus X) \rho_{\psi_j} T_{\psi_j}^{\widehat{A} \dag}(\Re \setminus X)]} \ ,
\end{equation}
\emph{and probabilities $p'_{1F}$, $p'_{2F}$, \ldots, respectively, with}
\begin{equation}
p'_{j F}=p_j \frac{p_{S_j}^{t}((A_0, \Re \setminus X))}{p_{S}^{t}((A_0, \Re \setminus X))}= p_{j} \frac{Tr[\rho_{\psi_j} T_{\psi_j}^{\widehat{A}}(\Re \setminus X)]}{Tr \Big [\sum_{j} p_{j} \rho_{\psi_j} T_{\psi_j}^{\widehat{A}}(\Re \setminus X) \Big ]} \ ,
\end{equation}
\emph{hence $S'_{F}$ is represented by the family of density operators} 
\begin{equation} \label{family_GLP_no}
\{ \rho_{S'_F}(H) \}_{H \in {\mathcal F}}=\{\sum_{j} p'_{j F} \frac{p_{S'_j F}^{d}(H)}{p_{S'_F}^{d}(H)} \rho_{\psi'_{j F}} \}_{H \in {\mathcal F}} \ .
\end{equation}

\vspace{.1cm}

It is then interesting to observe that the standard representation of $S_{F}$ is provided by the density operator
\begin{equation} \label{GLP_yes}
\rho_{S_{F}}=\sum_{j} p_{j F}\rho_{\psi_{j F}}=\sum_{j} p_j \frac{Tr[\rho_{\psi_j} T_{\psi_j}^{\widehat{A}}(X)]}{Tr \Big [\sum_{j} p_j \rho_{\psi_j}T_{\psi_j}^{\widehat{A}}(X) \Big ]} \frac{T_{\psi_j}^{\widehat{A}}(X)\rho_{\psi_j}T_{\psi_j}^{\widehat{A} \dag}(X)}{Tr[T_{\psi_j}^{\widehat{A}}(X)\rho_{\psi_j} T_{\psi_j}^{\widehat{A} \dag}(X)]} \ ,  
\end{equation}
while the standard representation of $S'_{F}$ can be obtained by replacing $p_{j F}$ with $p'_{j F}$, $\rho_{\psi_{j F}}$ with $\rho_{\psi'_{j F}}$ and $X$ with $\Re \setminus X$ in Eq. (\ref{GLP_yes}).

Eqs. (\ref{GLP'_yes}) and (\ref{GLP'_no}) coincide with Eqs. (\ref{genpost_dis_W}) and (\ref{projpostulate_W_no}), respectively, if $S$ is a pure state. Furthermore, Eq. (\ref{GLP_yes}) reduces to the L\"{u}ders formula whenever all detection probabilities coincide with 1, and also in this sense GLP generalizes the L\"{u}ders postulate. It must be stressed, however, that if a new property $H \in {\mathcal F}$ is measured on a physical object $x$ in the state $S_F$, one cannot use $\rho_{S_F}$ to evaluate the probabilities $p_{S_F}(H)$ and $p_{S_F}^{t}(H)$: one must calculate instead $\rho_{S_F}(H)$ and then apply Eqs. (\ref{AX_mixed_1}) and (\ref{conferma}), respectively, substituting $S$ with $S_F$ and $F$ with $H$. Similar remarks hold if $S'_{F}$ is considered in place of $S_F$. Notwithstanding this, it can be useful to refer also to the standard representation of mixtures in the ESR model, as we did in Sec. \ref{ignorance} and in the previous sections, in particular when dealing with the time evolution of states (Sect. \ref{genprojpost_measproc}).

\section{A dynamical justification of GPP\label{justification}}
We intend to show in this section that GPP can be partially justified by introducing a reasonable physical assumption on the evolution of the compound system made up of the (microscopic) physical object plus the (macroscopic) measuring apparatus. For the sake of simplicity and intuitivity we consider here only the discrete case, but the extension of our reasonings to the general case is straightforward.

Let $A$ be a discrete observable of QM represented by the self--adjoint operator $\widehat{A}$ and let $A_0$ be the generalized observable obtained from $A$. Whenever $A_0$ is measured on a physical object $x$ in the pure state $S$ represented by the unit vector $|\psi\rangle$, a natural extension of GPP consists in assuming that, if the outcome $a_n$ is obtained and the measurement is idealized and nondestructive, the final state of $x$ is given by Eq. (\ref{caso_particolare_psi_uno}). Hence, if the measurement is \emph{nonselective} (\emph{i.e.}, the outcome of the measurement remains unknown), the final state of $x$ is a mixture $\tilde{S}$ of the pure states $S_{F_0}$, $S_{F_1}$, $S_{F_2}$, \ldots represented by the density operators $|\psi_{F_0}\rangle\langle \psi_{F_0}|$, $|\psi_{F_1}\rangle\langle \psi_{F_1}|$, $|\psi_{F_2}\rangle\langle \psi_{F_2}|$, \ldots, with probabilities $p_{S}^{t}(F_0)$, $p_{S}^{t}(F_1)$, $p_{S}^{t}(F_2)$, \ldots, respectively, where the unit vectors $|\psi_{F_0}\rangle$, $|\psi_{F_1}\rangle$, $|\psi_{F_2}\rangle$, \ldots are given by Eq. (\ref{caso_particolare_psi_uno}) and the probabilities $p_{S}^{t}(F_0)$, $p_{S}^{t}(F_1)$, $p_{S}^{t}(F_2)$, \ldots, by Eq. (\ref{caso_particolare_n}). Hence the representation of $\tilde{S}$ in the ESR model is provided by the family
\begin{equation} \label{family_GLP_n}
\{\sum_{n \in {\mathbb N}_{0}} p_{S}^{t}(F_n) \frac{p_{S_{F_n}}^{d}(F)}{p_{S}^{d}(F)} |\psi_{F_n}\rangle\langle \psi_{F_n}| \}_{F \in {\mathcal F}} ,
\end{equation}
while the standard representation of $\tilde{S}$ is provided by the density operator
\begin{equation} \label{nonsel_gpp_explicit}
\tilde{\rho}=\sum_{n \in {\mathbb N}_{0}}p_{S}^{t}(F_n)|\psi_{F_n}\rangle\langle \psi_{F_n}| =p_{S}^{t}(F_0) |\psi_{F_0}\rangle\langle \psi_{F_0}|+\sum_{n \in {\mathbb N}} p_{\psi n}^{d}(\widehat{A})P_{n}^{\widehat{A}} |\psi\rangle\langle\psi| P_{n}^{\widehat{A}} \ .
\end{equation}
Let us denote by $g_1, g_2, \ldots$ the dimensions of the subspaces ${\mathscr S}_1, {\mathscr S}_2, \ldots$ associated with the eigenvalues $a_1, a_2, \ldots$, respectively, of $\widehat{A}$. Then, for every $n \in \mathbb N$, $P_{n}^{\widehat{A}}=\sum_{\mu} |a_n^{\mu}\rangle\langle a_{n}^{\mu}|$, where $\mu=1, \ldots, g_n$ and $ \{ |a_n^{\mu}\rangle \}_{\mu=1, \ldots,g_n}$ is an orthonormal basis in ${\mathscr S}_n$. Putting $|\psi\rangle=\sum_{n \in {\mathbb N}} \sum_{\mu} c_n^{\mu} | a_n^{\mu} \rangle$, Eq. (\ref{nonsel_gpp_explicit}) yields
\begin{equation} \label{entstate_W}
\tilde{\rho}=p_{S}^{t}(F_0) |\psi_{F_0}\rangle\langle\psi_{F_0}|+\sum_{n \in {\mathbb N}} p_{\psi n}^{d}(\widehat{A}) \sum_{\mu, \nu} c_n^{\mu} c_n^{\nu *} | a_n^{\mu} \rangle\langle a_n^{\nu} | .
\end{equation}

Let us now consider the macroscopic apparatus measuring $A_0$ as an individual example of a physical system $\Omega_M$ associated with the Hilbert space ${\mathscr H}_{M}$. Let $|0 \rangle, |1 \rangle, |2 \rangle, \ldots$ be the unit vectors of ${\mathscr H}_{M}$ representing the macroscopic states of $\Omega_M$ which correspond to the outcomes $a_0, a_1, a_2, \ldots$, respectively (hence $| 0 \rangle$ represents the macroscopic state of the apparatus when it is ready to perform a measurement or when the physical object $x$ is not detected), and let us assume that $\{| 0 \rangle, |1 \rangle, |2 \rangle, \ldots \}$ is an orthonormal basis in ${\mathscr H}_{M}$. Let $S_0$ be the initial state of the compound system made up of the physical object $x$ plus the macroscopic apparatus, represented by the unit vector $|\Psi_0\rangle=|\psi\rangle |0\rangle$. Because of the interpretation of the $a_0$ outcome provided in Sect. \ref{generalizedformalism}, the time evolution of the compound system must be such that the term $|\psi_{F_0}\rangle |0\rangle$ occurs in the expression of the unit vector $|\Psi\rangle$ representing the final state of the system in such a way that the probability of the $a_0$ outcome is $p_{S}^{t}(F_0)$. This makes it reasonable to suppose that the compound system undergoes the (generally \emph{nonlinear}, hence \emph{nonunitary}) time evolution
\begin{equation} \label{evolution}
|\Psi_0\rangle=|\psi\rangle |0\rangle =\sum_{n \in {\mathbb N}} \sum_{\mu} c_n^{\mu} | a_n^{\mu} \rangle |0\rangle
\longrightarrow
 |\Psi\rangle= \sum_{n \in {\mathbb N}} \alpha_{\psi n}  \sum_{\mu} c_n^{\mu} | a_n^{\mu} \rangle | n \rangle + \beta_{\psi 0} |\psi_{F_0}\rangle |0\rangle ,
\end{equation}
with $\alpha_{\psi n}=\sqrt{p_{\psi n}^{d}(\widehat{A})}e^{i \theta_{\psi n}}$ and $\beta_{\psi 0}=\sqrt{p_{S}^{t}(F_0)}e^{i \varphi_{\psi 0}}$ ($\theta_{\psi n}, \varphi_{\psi 0} \in \Re$), hence $\langle\Psi|\Psi\rangle= \sum_{n \in {\mathbb N}} |\alpha_{\psi n}|^{2} \sum_{\mu} |c_n^{\mu}|^{2} +|\beta_{\psi 0}|^{2}=1$ because of Eq. (\ref{caso_particolare_n}).

Let us perform now the partial trace with respect to ${\mathscr H}_{M}$ of the density operator $\rho_{C}=|\Psi\rangle\langle\Psi|$ representing the final state of the compound system after the interaction. We obtain
\begin{equation} \label{partrac}
Tr_{M} \rho_{C}=
p_{S}^{t}(F_0) |\psi_{F_0}\rangle\langle\psi_{F_0}|+\sum_{n \in {\mathbb N}} p_{\psi n}^{d}(\widehat{A}) \sum_{\mu, \nu} c_n^{\mu} c_n^{\nu *} | a_n^{\mu} \rangle\langle a_n^{\nu} | .
\end{equation}
By comparing Eqs. (\ref{entstate_W}) and (\ref{partrac}) we get
\begin{equation} \label{justif}
Tr_{M}\rho_{C}=\tilde{\rho} \ ,
\end{equation}
which provides a partial justification of GPP. This justification is not complete, because Eq. (\ref{justif}) does not imply that the state of $x$ is $\tilde{S}$, since the mapping $r$ in Eq. (\ref{diagram}) is not bijective, and $Tr_{M}\rho_{C}$ does not provide the representation that must be used to calculate probabilities of physical properties in the ESR model (Sect. \ref{gen_obse_mixed_case}). On the other side, if we recall that all probabilities are epistemic according to the ESR model (Sect. \ref{ignorance}) we see that the justification above does not introduce the problematic distinction between \emph{proper} and \emph{improper} mixtures that occurs in standard and unsharp QM, where the states obtained by performing partial traces are improper mixtures \cite{bs96,b98,gs07b}.

\end{document}